\documentclass{article}
\pdfoutput=1

\usepackage{arxiv}

\usepackage[utf8]{inputenc} 
\usepackage[T1]{fontenc}    
\usepackage{hyperref}       
\usepackage{url}            
\usepackage{booktabs}       
\usepackage{amsfonts}       
\usepackage{nicefrac}       
\usepackage{microtype}      
\usepackage{graphicx}
\usepackage[style=nature]{biblatex}
\usepackage{doi}
\usepackage{siunitx}
\usepackage{multicol}
\usepackage{enumitem}
\usepackage{float}
\usepackage{caption}
\usepackage{amsmath}
\usepackage{amssymb}

\DeclareSIUnit\rpm{rpm}
\DeclareSIUnit\fps{fps}
\DeclareSIUnit\cSt{cSt}
\DeclareSIUnit\atm{atm}

\title{Detection of nanobubbles between two liquid layers using atomic force microscopy meniscus force-distance measurements}

\author{ \href{http://orcid.org/0000-0001-8646-7661}{\includegraphics[scale=0.06]{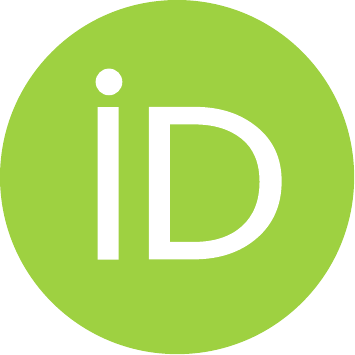}\hspace{1mm}Sam Peppou-Chapman}\\
	School of Chemistry\\
	The University of Sydney\\
	NSW 2006 \\
	\texttt{samuel.peppou-chapman@sydney.edu.au} \\
	\And
	\href{http://orcid.org/0000-0002-2174-8291}{\includegraphics[scale=0.06]{orcid.pdf}\hspace{1mm}Christopher Vega-S\'{a}nchez}\\
	School of Chemistry, The University of Sydney, NSW 2006 \\
	School of Electromechanical Engineering, \\Costa Rica Institute of Technology, Costa Rica \\
	\texttt{cveg6911@sydney.edu.au} \\
	\And
	\href{http://orcid.org/0000-0001-6058-0885}{\includegraphics[scale=0.06]{orcid.pdf}\hspace{1mm}Chiara Neto}\thanks{To whom correspondence should be directed.}  \\
	School of Chemistry\\
	The University of Sydney\\
	NSW 2006 \\
	\texttt{chiara.neto@sydney.edu.au} \\
}

\renewcommand{\shorttitle}{Detection of nanobubbles on LIS using AFM meniscus force measurements}

\hypersetup{
pdftitle={Detection of nanobubbles between two liquid layers using atomic force microscopy meniscus force-distance measurements},
pdfsubject={q-bio.NC, q-bio.QM},
pdfauthor={S. Peppou-Chapman, C. Neto},
pdfkeywords={Nanobubbles, AFM, LIS},
}

\begin{document}
\maketitle

\begin{abstract}
    The presence of nanobubbles on lubricant-infused surfaces (LIS) has so far been overlooked due to the difficulty in detecting them in such a complex system. We recently showed that anomalously large interfacial slip measured on LIS is explained by the presence of nanobubbles. \cite{Vega2021} Crucial to drawing this conclusion was the use of atomic force microscopy (AFM) force-distance spectroscopy to directly image nanobubbles on LIS. This technique provided vital direct evidence of the spontaneous nucleation of nanobubbles on lubricant-infused hydrophobic surfaces. In this paper, we describe in detail the data collection and analysis of AFM meniscus force measurements on LIS and show how these powerful measurements can quantify both the thickness and distribution of multiple coexisting fluid layers (\textit{i.e.} gas and oil) over a nanostructured surface.
\end{abstract}

\keywords{nanobubbles \and atomic force microscopy \and force-distance spectroscopy \and lubricant-infused surface \and liquid-infused surfaces \and SLIPS \and meniscus force measurement \and underwater measurements}

\begin{multicols}{2}
\section{Introduction}

Gas layers dramatically affect the flow boundary conditions in microfluidic systems, reducing drag by up to 75\%, \cite{lee2016superhydrophobic} but are frequently overlooked when one or more dimension is nanoscale due to difficulty in detecting them. Nanoscale gas layers, also known as surface nanobubbles, are extremely difficult to observe and characterize as they
are too small to be quantitatively analysed using optical
techniques which are typically used to study liquid-gas
interfaces. \cite{Alheshibri_Qian_Jehannin_Craig_2016}  \par

Surface nanobubbles were first reported in the year 2000 \cite{Alheshibri_Qian_Jehannin_Craig_2016, Ishida_Inoue_Miyahara_Higashitani_2000} and have been controversial ever since. The high Laplace pressure inside bubbles with radius of curvature smaller than \SI{100}{\nano\meter} (\textit{e.g.} $\Delta$P $\approx$ \SI{29}{\atm} for r = \SI{50}{\nano\meter} and $\gamma = \SI{72}{\milli\newton\per\meter}$) indicates that they should have very short lifetimes (\textit{i.e.} on the order of \SI{100}{\micro\second} \cite{Lohse_Zhang_2015}), however they are routinely seen to be stable on much longer timescales. \cite{Zhang_Maeda_Craig_2006} Their unexpected stability is due to the fact that, despite having thickness in the range of a few tens to hundreds of nanometers, they have micrometric lateral size, which produces a flat interfacial shape with low radius of curvature. This reduces the internal Laplace pressure and allows the bubble to remain stable on the immersed surface for hours to days. For a complete review of the field, the reader is directed to the review by Lohse and Zhang. \cite{Lohse_Zhang_2015} \par

Here, we report a method to detect and map the presence of gas layers on structured hydrophobic surfaces covered with a thin layer of a hydrophobic oil. This type of surface is known as lubricant-infused surface (LIS) and has been the topic of intense research over the past decade due to their desirable properties introduced by the presence of the entrapped lubricant layer, \cite{Peppou-Chapman_Hong_Waterhouse_Neto_2020} such as anti-fouling, \cite{Epstein_Wong_Belisle_Boggs_Aizenberg_2012, Sunny_Vogel_Howell_Vu_Aizenberg_2014, Maccallum_Howell_Kim_Sun_Friedlander_Ranisau_Ahanotu_Lin_Vena_Hatton_2015, Sotiri_Overton_Waterhouse_Howell_2016, Ban_Lee_Choi_Li_Jun_2017, Al-Sharafi_Yilbas_Ali_2017, Ware_Smith-Palmer_Peppou-Chapman_Scarratt_Humphries_Balzer_Neto_2018, Wang_Zhao_Wu_Wang_Wu_Xue_2019} anti-icing,  \cite{Kim_Wong_Alvarenga_Kreder_Adorno-Martinez_Aizenberg_2012, Kreder_Alvarenga_Kim_Aizenberg_2016, Subramanyam_Rykaczewski_Varanasi_2013, Yamazaki_Tenjimbayashi_Manabe_Moriya_Nakamura_Nakamura_Matsubayashi_Tsuge_Shiratori_2019} condensation enhancement \cite{Anand_Rykaczewski_Subramanyam_Beysens_Varanasi_2015, Al-Sharafi_Yilbas_Ali_2017,  Preston_Lu_Song_Zhao_Wilke_Antao_Louis_Wang_2018, Sett_Sokalski_Boyina_Li_Rabbi_Auby_Foulkes_Mahvi_Barac_Bolton_2019} and drag reduction \cite{Solomon_Khalil_Varanasi_2014, Kim_Rothstein_2016, Rosenberg_Van_Buren_Fu_Smits_2016, Wang_Zhang_Liu_Zhou_2016, Fu_Arenas_Leonardi_Hultmark_2017, Asmolov_Nizkaya_Vinogradova_2018,  Garcia-Cartagena_Arenas_An_Leonardi_2019,  Lee_Kim_Choi_Yoon_Seo_2019}.  \par

Generally, the \textit{L} in LIS is used interchangeably to indicate either \textit{liquid} \cite{Kim_Rothstein_2016} or \textit{lubricant}, \cite{Subramanyam_Rykaczewski_Varanasi_2013} as the most common liquids to impregnate surface structure are hydrophobic lubricants. \cite{Peppou-Chapman_Hong_Waterhouse_Neto_2020} Our recent insight showed that air and lubricant can both coexist within a hydrophobic surface structure and both act as lubricants, leading to drag reduction. \cite{Vega2021} Therefore, the distinction between \textit{liquid} and \textit{lubricant} is important as we showed that air is the fluid providing the greatest degree of lubrication when both are present. In this work, for clarity, the two lubricants will clearly be identified as oil (which could be any water-immiscible liquid lubricant) and as a gas layer. The term 'LIS' will be used to refer to a surface initially infused with a hydrophobic liquid lubricant before being submerged as we have done previously. \cite{Peppou_Chapman_AFM_Nanobubble_Mapping_2021,Peppou-Chapman_Hong_Waterhouse_Neto_2020} \par

Of particular interest is the ability of LIS to reduce interfacial drag. \cite{Schoenecker_Baier_Hardt_2014, Solomon_Khalil_Varanasi_2014, Kim_Rothstein_2016, Alinovi_Bottaro_2018, Ge_Holmgren_Kronbichler_Brandt_Kreiss_2018, Lee_Kim_Choi_Yoon_Seo_2019} Observed drag reduction \cite{Solomon_Khalil_Varanasi_2014, Kim_Rothstein_2016,Lee_Kim_Choi_Yoon_Seo_2019} is much higher than is expected by the interfacial slip model which predicts drag reduction only when the infused oil is less viscous than the flowing liquid. \cite{Vinogradova1999,Schoenecker_Baier_Hardt_2014,Alinovi_Bottaro_2018} Our recent work showed that the presence of isolated nanobubbles on silicone oil-infused Teflon wrinkled surfaces can quantitatively explain the observed drag reduction on LIS. \cite{Vega2021}

In this work we demonstrate that meniscus force mapping can be used to map hydrophobic oil and gas thickness simultaneously to reveal the pressence of nanobubbles on LIS. We describe AFM meniscus force measurements and how they can be used to detect and measure the thickness of a nanothin gas layer on top of a nanothin immiscible liquid layer, (\textit{i.e.} a nanobubble on a submerged LIS). To our knowledge, this is the first time two liquid/gas interfaces have been detected in a single AFM force-distance curve, and these force curves compiled to generate a time-resolved map of the spatial distribution of both phases. \par

\section{Materials and Methods}
    \subsection{Sample Preparation}
    Wrinkled Teflon surfaces were prepared as previously described \cite{Ware_Smith-Palmer_Peppou-Chapman_Scarratt_Humphries_Balzer_Neto_2018}. 
    Briefly, a shrinkable polystyrene substrate (Polyshrink\textsuperscript{TM}) was spin-coated with a thin layer ($\sim$\SI{40}{\nano\metre}) of Teflon AF (Chemours, 1.5\% in FC-40), and then annealed in an oven (France Etuves XFM020) at \SI{130}{\celsius}, inducing shrinking of the substrate and wrinkling of the top Teflon layer. \par
    The as-produced wrinkles were infused by pipetting an excess of the lubricant (approx. \SI{200}{\micro\litre\per\centi\metre\squared}) of silicone oil (\SI{10}{\cSt}, Aldrich), spreading it, and then depleting the oil through repeated immersion through an air/water interface \cite{Peppou-Chapman_Neto_2021} or using a spin coater. \cite{Peppou-Chapman_Neto_2018} \par  
    
    \subsubsection*{Control over air content in working fluids}
    Water with different air content was used in the experiments: Milli-Q water, used as produced, and gassed water. The procedure is described in \cite{Vega2021}. Briefly, the oxygen concentration in water was measured using a dissolved oxygen sensor (RCYACO, Model DO9100) and was used to estimate the air concentration in water. Milli-Q water as produced was air-saturated at atmospheric pressure (\SI{101}{\kilo\pascal}), and had an air content of $c_{air}\sim23.0\pm0.3$ \SI{}{\milli\gram\per\kilo\gram}. To produce gassed water, Milli-Q water was pressurized at \SI{203}{\kilo\pascal} to obtain an air content of $c_{air}\sim$ $44\pm4$ \SI{}{\milli\gram\per\kilo\gram}.
    
    \subsection{Meniscus Force Measurements}
    AFM meniscus force measurements were all performed using the force mapping feature on an MFP-3D (Asylum, Santa Clara, CA) using hydrophobized Multi-75 probes (k = \SIrange{1}{7}{\newton\per\meter}; Budget Sensors, Sofia, Bulgaria). The AFM probes are hydrophobized by depositing a thin layer of polydimethylsiloxane (PDMS) by chemical vapour deposition. The AFM probes are first cleaned using piranha solution, 3:1 sulfuric acid (98\%, Ajax) : hydrogen peroxide (30\%, Merck) for 5 minutes before being rinsed twice in Milli-Q water, once in toluene and dried under a gentle nitrogen flow. They are then placed in a glass staining jar with a small amount of uncured PDMS (Sylgard 184, Dow Corning) and placed in an oven at \SI{200}{\celsius} for 4 hours.  After cooling, they are rinsed once more with toluene and dried under a gentle nitrogen flow. The procedure deposits \SIrange{1}{2}{\nano\meter} of PDMS (by ellipsometry). \cite{Peppou-Chapman_Neto_2021} \par
    A custom-made sample holder is used to flood the samples with Milli-Q water \textit{in situ}. The cell consists of a superhydrophobic barrier with a small tubing through which water can be pumped, see our previous publication for details. \cite{Peppou-Chapman_Neto_2021} The custom cell was used for enhanced visibility and ease of use compared to the Asylum closed liquid cell when flooding a sample with water. This was important in previous work, but any underwater cell is sufficient to image nanobubbles using the technique described herein. \par
    All data was analysed using Python 3 \cite{CS-R9526} using packages included in the Anaconda scientific computing distribution \cite{anaconda}. Raw force curve data was exported to ASCII format before being loaded in the Python script and analysed using the algorithm described herein. All code used in this work is available online. \cite{Peppou_Chapman_AFM_Nanobubble_Mapping_2021}
    
\section{Results and Discussion}

\begin{figure}[H]
    \centering
    \includegraphics[width=0.9\linewidth]{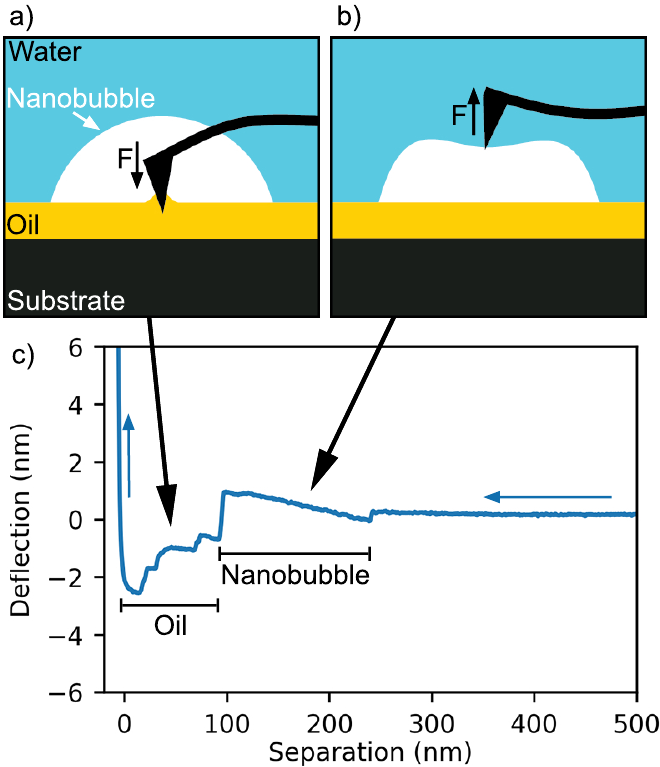}
    \caption{Schematic showing an AFM tip approaching a nanobubble over a layer of oil: both the (a) oil thickness and (b) the nanobubble thickness (not to scale) are measured due to the opposite forces they exert on the AFM tip. c) Example experimentally obtained force curve from which the nanobubble thickness and oil thickness are extracted.}
    \label{fig:Fig1_schematic}
\end{figure}

The detection of a thin gas layer sandwiched between the oil layer and the bulk water depends on detecting two features seen in AFM force-distance measurements. Oil layers are identified by a sharp negative deflection of the AFM cantilever, due meniscus formation (meniscus force measurements) \cite{Friedrich_Cappella_2020, Mate_Lorenz_Novotny_1990, Mate_Lorenz_Novotny_1989, Ally2010, Peppou-Chapman_Neto_2018, Scarratt_Zhu_Neto_2019} and gas layers are identified through the positive deflection of the cantilever. \cite{An_Tan_Ohl_2016, Tyrrell_Attard_2001, Walczyk_Schoenherr_2014a, Walczyk_Schoenherr_2014b, Wang_Zhao_Hu_Wang_Tai_Gao_Zhang_2017} Both these features are seen in the example force curve in \autoref{fig:Fig1_schematic}c, which was captured over an area identified as having both layers. Here, a positive deflection (\textit{i.e.} a force pushing the AFM tip away from the surface) is seen at larger separations and a negative deflection (\textit{i.e.} a force drawing the tip towards the surface) is seen a smaller separations, corresponding to the nanobubble and oil layer, respectively. \par

In this work, only extension curves were used and are presented, as the forces of interest occur as the tip moves towards the surface. Unless specified otherwise, all data was collected on a LIS composed of wrinkled Teflon infused with \SI{10}{\cSt} silicone oil. \cite{Ware_Smith-Palmer_Peppou-Chapman_Scarratt_Humphries_Balzer_Neto_2018, Vega2021} \par

    \subsection{Meniscus Force Measurements}
    Meniscus force measurements are a subset of AFM force spectroscopy in which the dominant force on the cantilever is due to the formation of a liquid meniscus around the AFM tip. In contrast to the case where the AFM tip first encounters the solid surface (see \autoref{fig:Fig2_ex_FCs}a), when the AFM tip first contacts the air/liquid interface, a liquid meniscus forms around the tip and pulls down the cantilever towards the surface (see \autoref{fig:Fig1_schematic}a), causing a rapid negative deflection (see \autoref{fig:Fig2_ex_FCs}b). The distance between this so-called `jump-in' and the position where the AFM tip contacts the hard substrate underneath is the thickness of the liquid film, and can be revealed with nanoscale accuracy. \cite{Friedrich_Cappella_2020, Mate_Lorenz_Novotny_1990, Mate_Lorenz_Novotny_1989, Ally2010, Peppou-Chapman_Neto_2018, Scarratt_Zhu_Neto_2019} A detailed discussion on the accuracy of the technique is provided \cite{Friedrich_Cappella_2020, Mate_Lorenz_Novotny_1990, Mate_Lorenz_Novotny_1989}. \par 
    An advantage of this technique is that it is agnostic to the identity of the two phases. As long as the interfacial energy of the liquid is sufficiently high to observe a jump-in when the AFM tip touches it, the interface will be detected in the force curve. As a result, we have used this technique to map the thickness of a hydrophobic oil layer both in air \cite{Peppou-Chapman_Neto_2018,Tonelli_Peppou-Chapman_Ridi_Neto_2019} and underwater \cite{Peppou-Chapman_Neto_2021} as the water/oil interface has sufficient wetting contrast to produce the required jump-in. \autoref{fig:Fig2_ex_FCs}b shows the shape of a typical force curve of a thin oil film underwater where a sharp negative deflection is seen at about \SI{50}{\nano\meter} separation, before a sharp positive deflection when the tip contacts the underlying substrate. 

    \subsection{Force-Distance Curves on Surface Nanobubbles}
    Due to their nanoscale dimensions, nanobubbles are not routinely studied using optical techniques typical in the study of larger gas bubbles. Instead, the nanoscale nature of AFM measurements has made it the technique of choice for the study of nanobubbles using both tapping mode and force-distance spectroscopy. \cite{Lohse_Zhang_2015,Alheshibri_Qian_Jehannin_Craig_2016}  \par
    
    In AFM force-distance measurements, nanobubbles are identified through a characteristic positive deflection (pushing the tip away from the substrate, see \autoref{fig:Fig1_schematic}b) due to deformation of the air/water interface. \cite{An_Tan_Ohl_2016, Walczyk_Schoenherr_2014b} \autoref{fig:Fig2_ex_FCs}c shows the shape of a typical force curve taken on a nanobubble, showing a section of force curve with a positive gradient between the zero force baseline and the hard contact point. Many publications have confirmed that this particular feature in the force curve shape is due to nanobubbles. \cite{An_Tan_Ohl_2016,Tyrrell_Attard_2001, Walczyk_Schoenherr_2014a, Walczyk_Schoenherr_2014b, Zhang_Maeda_Craig_2006, Wang_Zhao_Hu_Wang_Tai_Gao_Zhang_2017} \par

    \subsection{Imaging Nanobubbles on LIS}
\begin{figure}[H]
        \centering
        \includegraphics[width=\linewidth]{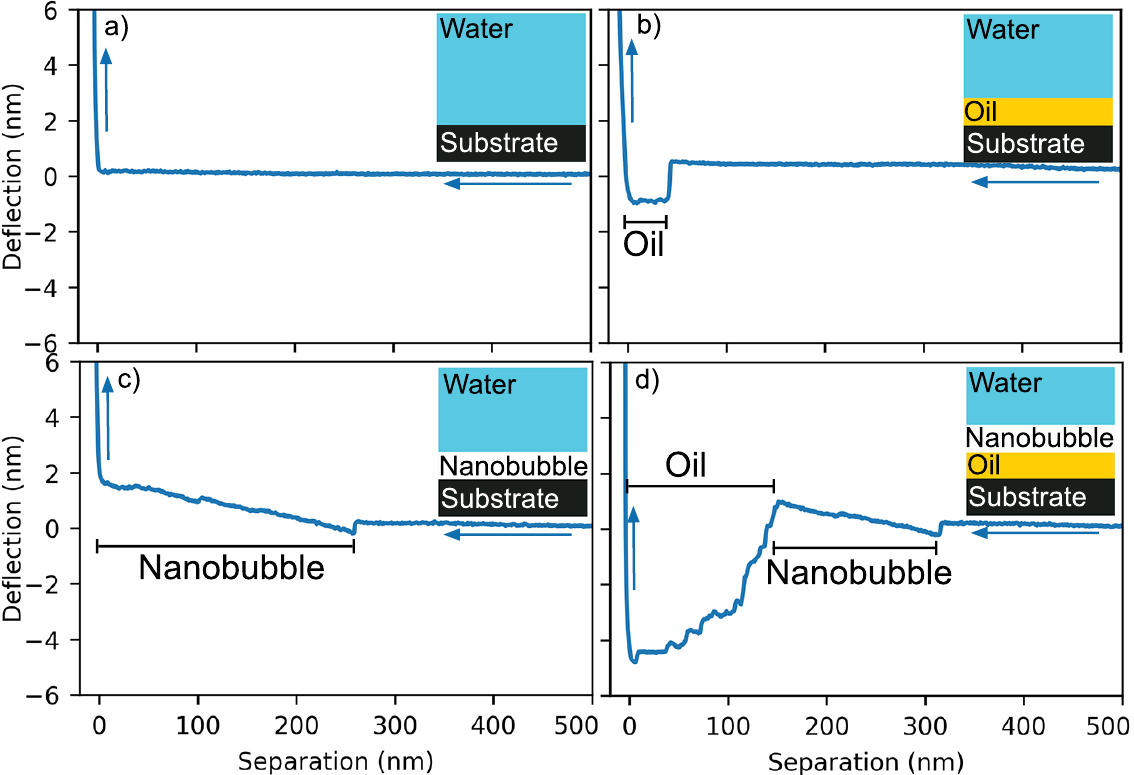}
        \caption{Example of the four types of force curves observed on silicone oil-infused Teflon LIS under Milli-Q water. Extension force curve as the AFM tip comes in contact with a) bare solid substrate under water; b) an oil film on the solid substrate under water; c) a gas layer on the solid substrate under water; d) a gas layer on an oil layer on the solid substrate under water.}
        \label{fig:Fig2_ex_FCs}
    \end{figure}

    The two types of force curves described above combine when a gas layer is present on top of an oil layer underwater. In its approach towards the surface, the tip first deflects away due the deformation of the bubble and then towards the substrate due to meniscus formation when it contacts the oil layer, see \autoref{fig:Fig2_ex_FCs}d. The first point at which the positive deflection occurs is used to indicate the top of the gas layer and the first point of the negative deflection indicates the top of the oil layer. This interpretation may lead to small under- or overestimation of the layer thickness in certain cases. These effects are explored more in Section \ref{sec:limitations}. \par
    
    The process of imaging nanobubbles on LIS is similar to that of imaging the oil layer underwater, \cite{Peppou-Chapman_Neto_2021} with most of the difference consisting in the data analysis, see Section \ref{sec:data_anaysis}.  \par
    The four expected configurations of fluid layers and example force curves for an immersed LIS under static conditions are shown in \autoref{fig:Fig2_ex_FCs}: water on substrate, water on oil on substrate, water on gas on substrate, water on gas on oil on substrate. We do not observe a thick layer of oil on a nanobubble, and indeed a thick layer is not expected to be stable as the static pressure and Laplace pressure would destabilise the oil layer. \cite{Kreder_Daniel_Tetreault_Cao_Lemaire_Timonen_Aizenberg_2018} On the other hand, a nano-thin film ($\lesssim$ \SI{5}{\nano\meter}) of oil is expected to cloak the nanobubble due to the positive spreading parameter of oil at the water/air interface, but was not detectable with the current experimental set-up (see Section \ref{sec:limitations} for discussion of possible reasons). \par
    
    The time required for the formation of nanobubbles is dependent on the gas content in the water. Nanobubbles appeared faster in water gassed with air (\textit{e.g.} for water air content $c_{air}\sim44\pm4$ \SI{}{\milli\gram\per\kilo\gram} they appeared instanteously, see \autoref{fig:Fig6_maps_cross_sec}e), while they appeared a few hours after immersion in plain Milli-Q water, which was saturated at atmospheric pressure ($c_{air}\sim23.0\pm0.3$ \SI{}{\milli\gram\per\kilo\gram}). \autoref{fig:Fig6_maps_cross_sec}a-d show the nucleation of a nanobubble on a surface through successive mapping of the same location. Each map took $\sim$35 minutes and the bubble appeared between successive maps. The surface was submerged for a total 2-3 hours with static conditions before the nanobubble appeared. \par
    
    \subsection{AFM Probe Considerations}
    \label{sec:AFM_probe}
    The choice of tip shape and chemistry is important to achieving a clear image of both the oil and the gas layer. In AFM mapping techniques, a sharp tip is generally preferred as it gives maximum spatial resolution, but for meniscus force measurement a thinner tip decreases the force on the cantilever as the force is determined by the length of the contact line. \cite{Cappella_2017,Friedrich_Cappella_2020} As a result, there is a balance between spatial resolution and force resolution. If a tip of a known shape is used (\textit{e.g.} cylindrical), the length of the contact line is known and the surface tension of the fluid can be calculated. \cite{McGuiggan_Wallace_2006} \par
    Similarly, cantilever spring constant is critical to successful meniscus force measurements. A low spring constant is needed for the cantilever to deflect due to meniscus formation and to deflect from the reaction force caused by nanobubble deformation. Additionally, a lower spring constant allows for a sharper tip to be used. However, the spring constant cannot be too low (\textit{e.g.} a contact mode probe), as the cantilever is not able to break free from the meniscus during retraction. Here, force modulation AFM probes with a spring constant of $\sim$ \SI{5}{\newton\per\meter} and a sharp tip were used, as these provide sufficiently large deflection values with good spatial resolution.\par
    For imaging hydrophobic oil layers, tip chemistry determines both whether a meniscus is formed in the first place and the magnitude of the force exerted by this meniscus (as this is determined by the length of the contact line). The fluid layer that forms the meniscus should preferentially wet the tip for a meniscus to form. In air, this is trivial as almost all fluids wet the high surface energy silicon nitride which makes up most AFM tips. Underwater, this is less simple as a hydrophilic tip may be wet preferentially by the water and therefore a clear jump-in from a hydrophobic oil might not be apparent. Here, the tip was hydrophobized with a thin layer of PDMS to ensure a wettability contrast between the water and the thin oil layer. An unmodified tip was seen to work initially, but the image quality deteriorated quickly, leading to the loss of jump-in. \par
    For imaging nanobubbles, the opposite is required. The thickness of a nanobubble is better imaged using a hydrophilic tip as the tip is less likely to penetrate into the bubble so all changes in deflection are due to nanobubble deformation. \cite{Walczyk_Schoenherr_2014b} A hydrophobic tip still shows deflection due to nanobubble deformation, however, it also shows signs of meniscus formation as the tip contacts the nanobubble. This aspect is discussed further in Section \ref{sec:limitations}.

    \subsection{Data Analysis}
    \label{sec:data_anaysis}
    The key to using AFM meniscus force measurements for mapping of multiple layers was the automated data analysis with feature recognition to determine the points at which the tip contacts different interfaces. This section describes the logic used by our Python script to calculate the thickness of both the gas and oil layer from the collected force curves.  \par
    
    \begin{figure*}
        \centering
        \includegraphics[width=\textwidth]{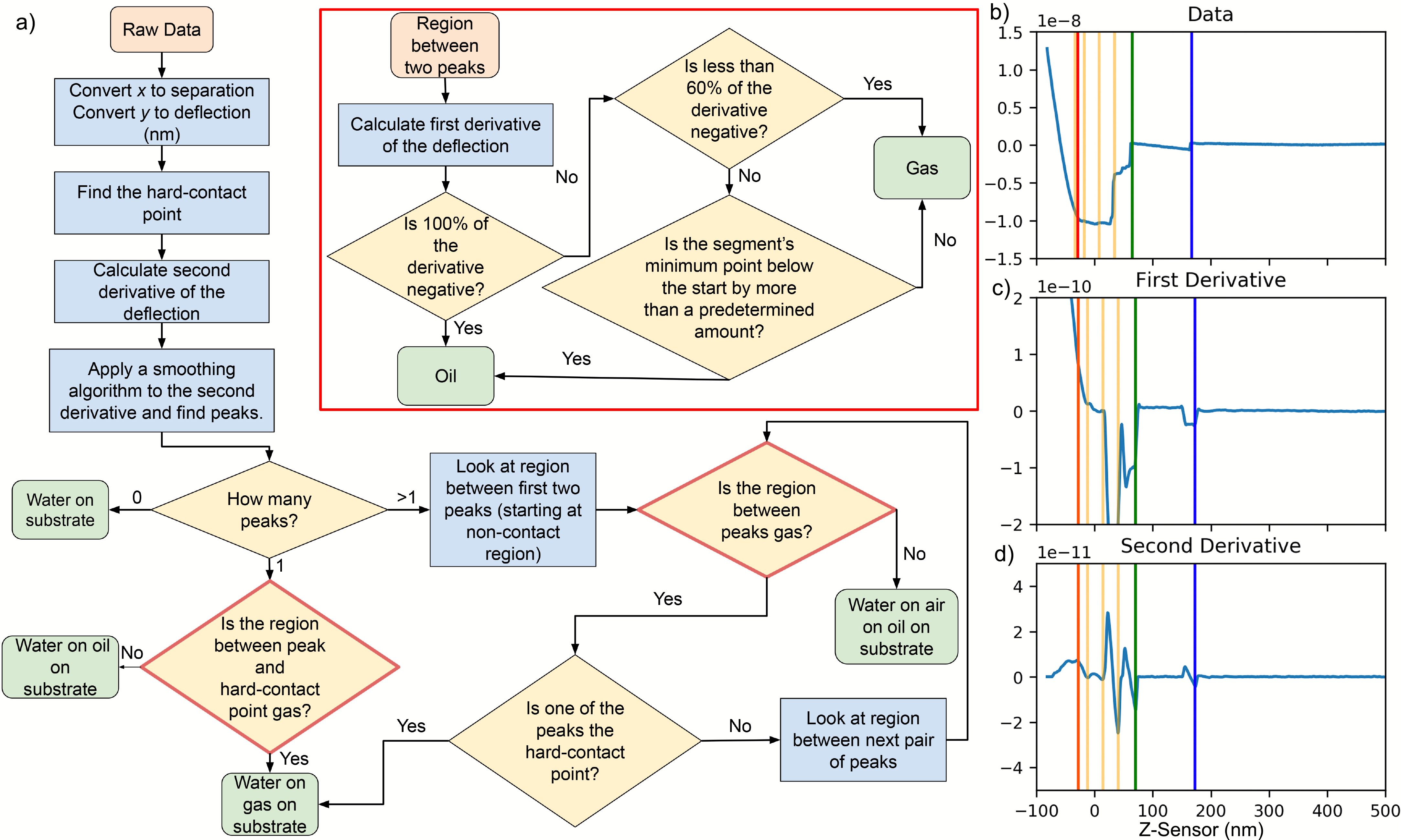}
        \caption{a) Flowchart showing the process used to characterise each force curve in the raw force map data. The inset in the red box is the logic used in the decision nodes with a red outline to determine whether the data between two features of a force curve is gas or lubricant. See GitHub \cite{Peppou_Chapman_AFM_Nanobubble_Mapping_2021} for code. b) Raw force curve data of cantilever deflection (y scale is in \SI{}{\meter}) as function of \textit{z}-displacement for an example force curve. c) The first derivative of b) (y scale is \SI{}{\meter\per\nano\meter}). d) The second derivative of b) (y scale is \SI{}{\meter\per\nano\meter\squared}). The orange vertical lines represent negative peaks picked from the second derivative. The red, green, and blue lines represent the location of the substrate, the top of the oil and the top of the nanobubble, respectively.}
        \label{fig:Fig3+4_flow_FC}
    \end{figure*}
    
    For an underwater LIS there are four possible scenarios for the fluid layers each force curve may encounter (see \autoref{fig:Fig2_ex_FCs}):
    \begin{enumerate}[label=(\alph*)]
        \item Water on substrate;
        \item Water on oil on substrate;
        \item Water on gas on substrate;
        \item Water on gas on oil on substrate;
    \end{enumerate}

    In our previous work, \cite{Peppou-Chapman_Neto_2018, Tonelli_Peppou-Chapman_Ridi_Neto_2019, Peppou-Chapman_Neto_2021} the script only detected an oil layer (\textit{i.e.} scenarios (a) and (b)), where only two key points needed to be detected: the jump-in and hard contact point. As a result of the added complexity added by the gas layer, a completely new detection algorithm was used here to detect all four of these scenarios. The features associated with the gas layer are less pronounced than in previous iterations, and so a lower threshold for detection of features needed to be implemented.\par
    Here, the algorithm utilises automated peak finding to find areas of rapid changes in the gradient of the deflection and the fact that the gradient of the deflection is opposite signs when the tip is in contact with a nanobubble versus when in contact with a meniscus due to an oil layer. The general procedure used in the script is (see also \autoref{fig:Fig3+4_flow_FC}a): 
    \begin{itemize}
        \item Convert piezo movement data to separation, and photo diode signal to deflection, using sensitivity of cantilever calculated from compliance region measured on a hard surface.
        \item Find the hard-contact point by looking for the first point where the gradient changes substantially while moving towards larger separations from the turn around point.
        \item Calculate the second derivative to the whole data and apply a smoothing algorithm to the y-data (deflection). 
        \item Use a peak finding algorithm to find negative peaks in the smoothed second derivative (orange lines in \autoref{fig:Fig3+4_flow_FC}b-d).
        \item If no peaks are detected, there is no oil or gas (\textit{i.e.} case in \autoref{fig:Fig2_ex_FCs}a)
        \item If one peak is detected, then the region between it and the hard contact point can either be oil or gas. Check if gas or oil using the algorithm outlined below. (\textit{i.e.} case in \autoref{fig:Fig2_ex_FCs}b,c).
        \item If multiple peaks are detected, each segment between two peaks is checked to see if it is gas or not using the algorithm outlined below. The first segment identified as oil demarcates the boundary between the gas and oil (\textit{i.e.} case in \autoref{fig:Fig2_ex_FCs}d). If no oil is detected before reaching the hard contact point, then there is no oil present and the layer is entirely gas (i.e case in \autoref{fig:Fig2_ex_FCs}c).
    \end{itemize}

    To determine whether a segment between two of the negative peaks identified in the second derivative is gas or oil, the following algorithm is used (see also inset within the red box in \autoref{fig:Fig3+4_flow_FC}a):
    \begin{itemize}
        \item If 100\% of the first derivative is negative – the portion is oil. This is because the force acting on the cantilever is proportional to the length of the contact line and this force always increases (due the triangular shape of the tip) as the tip moves towards the surface – causing the deflection to become more negative. 
        \item If more than 60\% of the points are negative and the end is lower than the start by a threshold deflection (\SI{0.5}{\nano\meter} in this work, determined using trial and error), then portion is considered oil. Otherwise, it is gas.
    \end{itemize}

    The script then outputs the height of the hard contact point (absolute height from $z$-sensor data) and the thickness of the oil or gas layers found (calculated as the distance between relevant peaks in separation). \autoref{fig:Fig5_example_map}a shows a map with example forces curves and their corresponding location in the map. The force curves (\autoref{fig:Fig5_example_map}b-g) also show the features located by the algorithm described above (blue = start of gas layer, green = start of oil layer, and red = start of the substrate). \par
    
        \subsubsection{Validation}
        
        This analysis was validated by a manual review of analysed force curves to judge if the script had picked the correct locations for the start of gas and oil layers. Force curves from multiple maps were plotted with the vertical lines showing the interfaces as picked by the script as in \autoref{fig:Fig3+4_flow_FC}b and \autoref{fig:Fig5_example_map} and judged by eye if these were correct. A total of 494 force curves were selected at random and 93\% of them were judged to be correctly fitted. Potential reasons for incorrect fitting are discussed below in Section \ref{sec:limitations}. \par

    \subsection{Data Visualisation}
    
\begin{figure}[H]
        \centering
        \includegraphics[width=\linewidth]{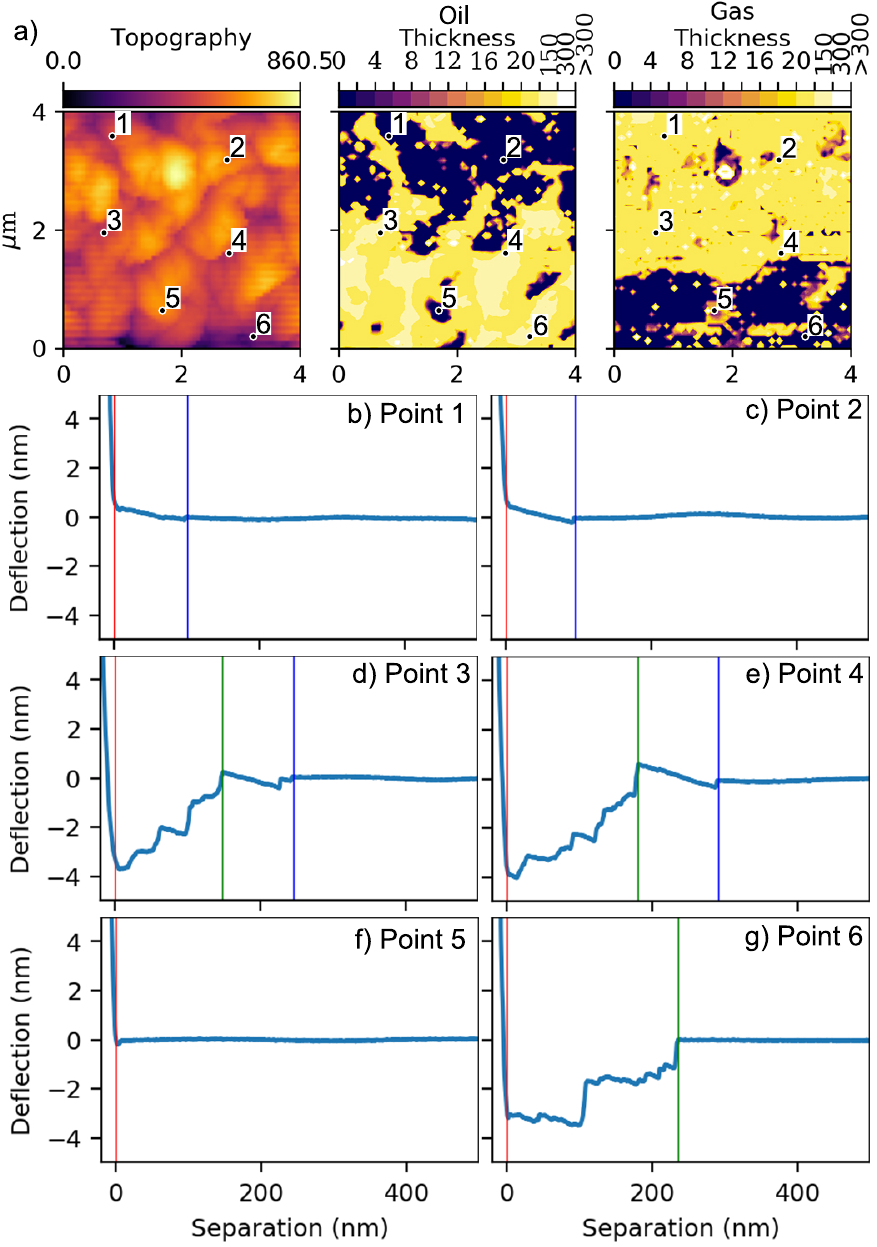}
        \captionof{figure}{a) Map of a nanobubble on a LIS with (b-g) example force curves shown for points with different fluid layers. The map (a) is presented with three panels showing the topography of the underlying Teflon wrinkles (left), the thickness of the oil (middle) and the thickness of the gas (left); the units of all colour scales is \SI{}{\nano\meter}. Points 1 and 2 show an example of a nanobubble directly on the substrate. Points 3 and 4 show an example of a nanobubble on oil. Point 5 shows an example of the water directly contacting the substrate and point 6 shows an example of water contacting the oil. The vertical lines in the force curves show where the script has found the start of the different fluid layers, blue = start of gas layer, green = start of oil layer, and red = start of the substrate.}
        \label{fig:Fig5_example_map}
   \end{figure}

    As shown in \autoref{fig:Fig5_example_map}, the data is presented as multi-panel maps showing the spatial distribution of the different quantities measured. First, the sample topography is presented (extracted from the hard contact point of the force curves). Then maps of the oil thickness and the gas thickness are presented separately. Both oil and gas thickness are presented as non-linear contours where colours are evenly spaced at thickness of \SI{2}{\nano\meter} up to \SI{20}{\nano\meter} and then contours of \SI{150}{\nano\meter}, \SI{300}{\nano\meter} and >\SI{300}{\nano\meter} to give an indication of distribution of both thin layers and thick layers. This method of presentation was selected as it gives the ability to easily attribute the effect of topography on the distribution of either the liquid or the gas layer. \par
    
    \begin{figure*}
        \centering
        \includegraphics[width=\textwidth]{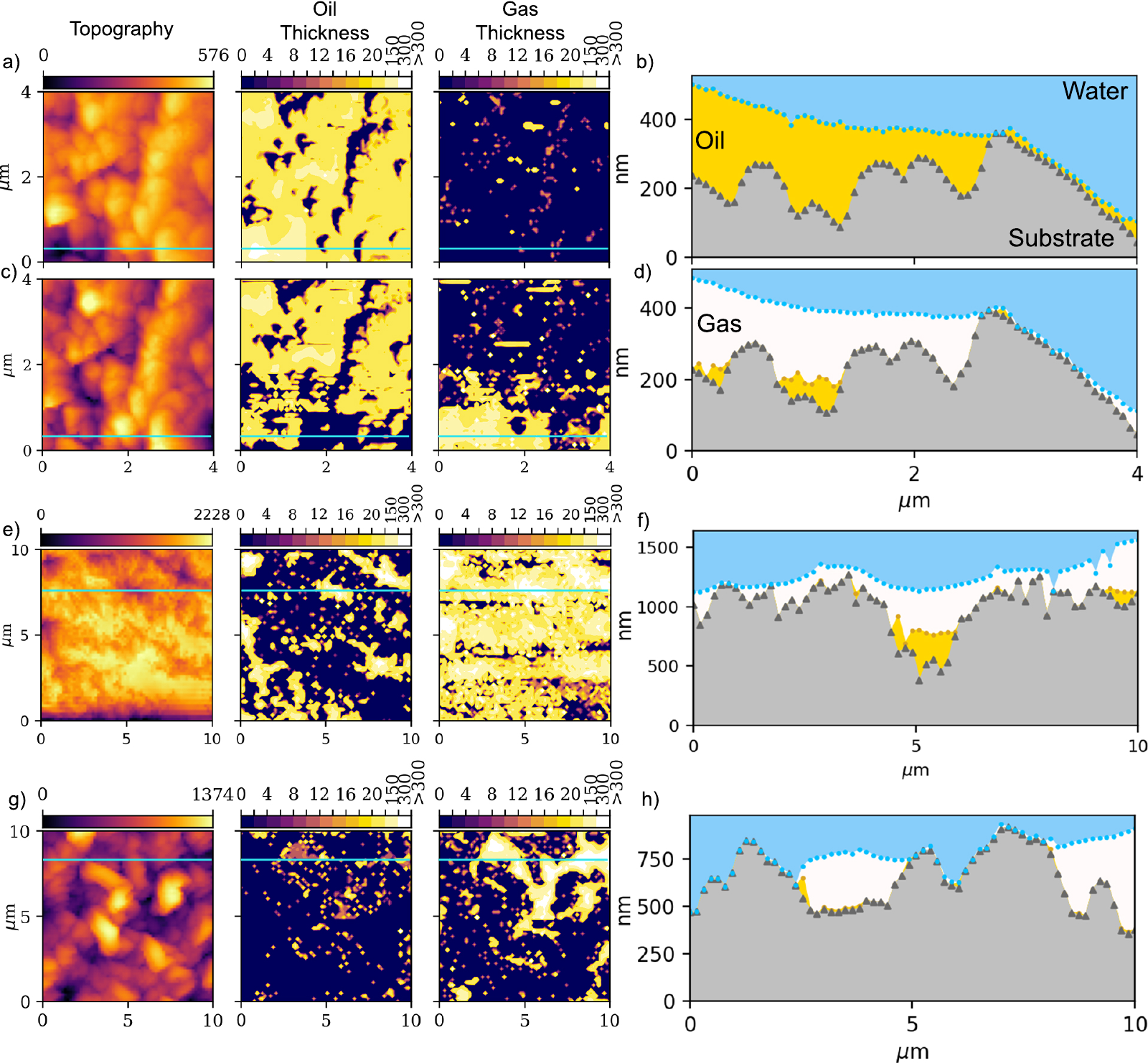}
        \caption{AFM force maps showing oil and gas thickness (a, c, e, g) with cross sections (b, d, f, h) corresponding the cyan lines in the maps. The units of all colour scales is \SI{}{\nano\meter}. The marker symbols in the cross section represent points at which the height was measured and colors were added to help visualise the different layers. a) and b) are successive maps on the same area of the surface, showing the appearance of a nanobubble \textit{in situ}. The force curves were collected and analysed using the same parameters. e, f)  example of a LIS submersed in gassed water ($c_{air}\sim44\pm4$ \SI{}{\milli\gram\per\kilo\gram}) imaged immediately after submersion. g, h) a superhydrophobic surface (the underlying TW substrate of the LIS without any silicone oil applied) showing a partially collapsed Cassie state. The non-zero measurements of oil thickness at multiple points on this map are a result of the feature detection algorithm not always being able to discern the difference between oil and gas or the particular force curve being noisy. See Section \ref{sec:limitations} for more information.}
        \label{fig:Fig6_maps_cross_sec}
    \end{figure*}

    Cross sectional profiles of these maps can be extracted to visualise the shape of the interfaces, as shown in \autoref{fig:Fig6_maps_cross_sec}, in which color was added to the line profiles to highlight the different fluid layers. Each cross section corresponds the cyan line in the adjacent maps. As mentioned above, part (a) and part (b) in \autoref{fig:Fig6_maps_cross_sec} are sequential maps (taken approximately 35 mins apart) of the same surface before and after a nanobubble nucleates. \autoref{fig:Fig6_maps_cross_sec}e,f shows a cross sections from the LIS in \autoref{fig:Fig5_example_map} while \autoref{fig:Fig6_maps_cross_sec}g,h shows gas pockets on an immersed superhydrophobic surface (Teflon wrinkles with no oil present) in a partially-collapsed Cassie state. Despite the apparently high resolution of the contact line of the mapped interfaces, the local contact angle values can not be estimated, as discussed in Section \ref{sec:limitations}.

    \subsection{Limitations and Sources of Error}
    \label{sec:limitations}
    
    There are several limitations and sources of error related to this technique which are somewhat accounted for by the mapping nature which allows individual errant pixels to be ignored. This section discusses limitations and sources of error in this technique. \par
    
    The accuracy of the measured thickness depends on many factors and is not uniform for all film thicknesses. Very thin (<\SI{5}{\nano\meter}) films of either gas or oil are difficult to both detect and distinguish. The technique’s ability to detect a nanoscale film is proportional to the film's thickness as thicker films (either gas or oil) produces a greater deflection while films of just a few nanometers produce a deflection on the same order as measurement noise. As a result, regions of zero film thickness may contain undetected films of a few nanometers. \par
\begin{figure}[H]
        \centering
        \includegraphics[width=\linewidth]{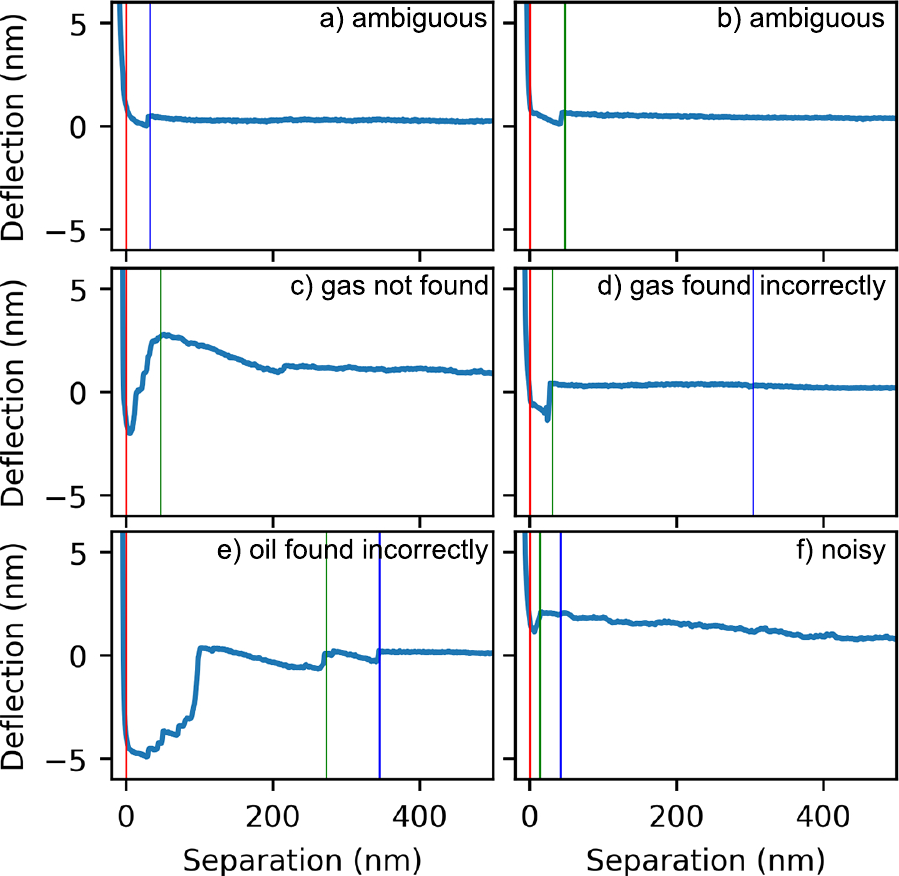}
        \caption{Force curves showing the limitations of the automatic feature detection script presented in this work. Vertical lines indicate where the analysis script found an interface with blue = start of gas layer, green = start of oil layer, and red = start of the substrate. a,b) Two force curves obtained on adjacent locations, where the identity of the feature is ambiguous, with the script identifying in (a) a gas layer and in (b) a oil layer; c) an example of the script not detecting gas; d) an example of the script incorrectly detecting the gas/water interface; e) an example of the script incorrectly detecting the oil/gas interface instead of more gas; f) an example of a noisy force curve, likely due to vibrations. }
        \label{fig:Fig7_bad_Fcs}
    \end{figure}

    Automated feature detection exacerbates this limitation as there is a trade-off between sensitivity (\textit{i.e.} detecting smaller deflections) and avoiding false detection at large separations (\textit{e.g.} more than \SI{100}{\nano\meter} from the first interface as in \autoref{fig:Fig7_bad_Fcs}d). The noise present in the oil and gas thickness maps in \autoref{fig:Fig5_example_map} is the result of imperfect fitting that occurs due to higher feature detection sensitivity. Using a thicker tip would reduce this effect but, as mentioned above, would reduce lateral resolution. Slightly thicker films (up to tens of \SI{}{\nano\meter}) can be detected but with fewer data points to judge whether the feature in the force curve is due to oil or gas, the script is prone to mislabelling these. \autoref{fig:Fig7_bad_Fcs}a,b shows examples of two features at small separations with deflections $\sim$\SI{1}{\nano\meter} which are ambiguous and were identified as different features by the script, despite being adjacent pixels in the map.  \par
    
    The presence of a nano-thin layer of oil spread on the AFM tip cannot be excluded, which would increase the measured thickness of the oil layer. As with our previous publications, this does not seem to be an issue as there are multiple cases where no oil thickness is detected either under a nanobubble or elsewhere (\textit{e.g.} \autoref{fig:Fig2_ex_FCs}a,c). However, the lack of sensitivity to very thin films (<\SI{5}{\nano\meter}) described in the previous paragraph may hide the effects of such a film. \par
    
    The thickness of the oil layer may be overestimated if long range van der Waals attraction between the tip and an interface draws the interface up to ‘meet’ the tip. \cite{Ally2010} In our previous publication, this effect was suppressed by sufficiently fast scan rate in the force curve \cite{Peppou-Chapman_Neto_2018}. As a result, here the fastest scan rate possible on our AFM (\SI{2}{\hertz}) was used to ensure this effect is minimized. This deformation will also effect the gas layer, with the air/water interface deforming to meet the tip, especially for a hydrophobic tip. \cite{Walczyk_Schoenherr_2014b} As a result of this deformation, Walczyk \& Sch{\"o}nherr \cite{Walczyk_Schoenherr_2014b} define the top of the a nanobubble measured with a hydrophobic tip to be where the force curve crosses zero deflection after the initial jump-in. Here potential long-range attraction of the air/water interface was ignored, due to the high scan rates used and because this correction would cause the calculated gas height to be heavily dependent on the quality of the baseline correction. Additionally, Walczyk \& Sch{\"o}nherr's definition of the top of a nanobubble assumes that the AFM tip only contacts the bubble with minimal contact line at zero deflection (\textit{i.e.} the only force on the cantilever is due to nanobubble deformation), which is impossible given the fact that a jump-in is seen, signifying meniscus formation and a non-trivial contact line. \par

    There are limitations of the experimental setup which contribute noise in the maps. Force curves collected under water are particularly sensitive to vibrations, as they can be transmitted through the liquid to the cantilever. This means that a higher proportion of force curves are noisy throughout their entire range of motion, compared to the same force measurement in air, leading to incorrect feature detection (see \autoref{fig:Fig7_bad_Fcs}f). \par
    
    The cross sections generated from the map cannot be used to quantify contact angle values as the interface is 3-dimensional and contributions from in-plane and out-of-plane features are impossible to account for. Additionally, the shape of the interface may be slightly deformed due van der Waals interactions not being consistent across the surface. \par

\section{Conclusion}
In summary, we have shown that a single AFM force-distance curve can capture the thickness of both a gas layer and an immiscible liquid layer underwater simultaneously. By mapping a surface with force curves and analysing the data automatically, the thickness of the gas and the immiscible liquid are both spatially resolved. This presents an exciting new technique to study systems such as the nucleation of nanobubbles on LIS. \cite{Vega2021} There are still many outstanding questions related to nanobubbles on LIS including the effects of substrate topography, the presence of surfactants or ion concentration and effect of static pressure. This force mapping technique will enable detailed studies of these parameters and could help in establishing whether nanobubbles are more stable on superhydrophobic or on LIS.

\printbibliography

\end{multicols}
\end{document}